\documentclass[12pt]{article}
\usepackage{amssymb,latexsym}
\oddsidemargin=\evensidemargin
\advance\topmargin by-\headheight
\def\exp{\mathop{\rm exp}\nolimits}
\def\FA{\mathop{\rm FACTOR}\nolimits}

\def\GCD{\mathop{\rm GCD}\nolimits}
\def\id{\mathop{\rm id}\nolimits}
\def\im{\mathop{\rm im}\nolimits}
\def\ker{\mathop{\rm ker}\nolimits}
\def\lg{\langle}
\def\MM{\mathop{\rm INVERSE}\nolimits}
\def\mapright#1{\smash{\mathop{\rightarrow}\limits^{#1}}}
\def\mod{\mathop{\rm mod}\nolimits}
\def\NN{{\mathbb N}}
\def\ov{\overline}

\def\R{{\cal R}}
\def\rg{\rangle}
\def\S{{\cal S}}
\def\sym{\mathop{\rm Sym}\nolimits}

\def\ZZ{{\mathbb Z}}

\def\proof{{\bf Proof}.\ }
\def\bull{\vrule height .9ex width .8ex depth -.1ex }
\def\subs{\stepcounter{subsection}{\bf\thesubsection{.}}
\addtocounter{subsection}{-1}\refstepcounter{subsection}}
\def\subsn{\vspace{2mm}\subs}

\newtheorem{formula}{}[section]

\newtheorem{definition}[formula]{Definition}
\newtheorem{corollary}[formula]{Corollary}
\newtheorem{remark}[formula]{Remark}
\newtheorem{lemma}[formula]{Lemma}
\newtheorem{theorem}[formula]{Theorem}

\def\thrm{\begin{theorem}}
\def\thrml#1{\begin{theorem}\label{#1}}
\def\ethrm{\end{theorem}}
\def\rmrk{\begin{remark}}
\def\rmrkl#1{\begin{remark}\label{#1}}
\def\ermrk{\end{remark}}
\def\dfntn{\begin{definition}}
\def\dfntnl#1{\begin{definition}\label{#1}}
\def\edfntn{\end{definition}}
\def\nmrt{\begin{enumerate}}
\def\enmrt{\end{enumerate}}
\def\tm#1{\item[{\rm (#1)}]}
\def\qtn{\begin{equation}}
\def\qtnl#1{\begin{equation}\label{#1}}
\def\eqtn{\end{equation}}
\def\lmm{\begin{lemma}}
\def\lmml#1{\begin{lemma}\label{#1}}
\def\elmm{\end{lemma}}
\def\crllr{\begin{corollary}}
\def\crllrl#1{\begin{corollary}\label{#1}}
\def\ecrllr{\end{corollary}}

\begin{document}
\title{Homomorphic public-key cryptosystems \\
and encrypting boolean circuits }
\author{
Dima Grigoriev \\[-1pt]
\small IRMAR, Universit\'e de Rennes \\[-3pt]
\small Beaulieu, 35042, Rennes, France\\[-3pt]
{\tt \small dima@maths.univ-rennes1.fr}\\[-3pt]
\small http://www.maths.univ-rennes1.fr/\~{}dima
\and
Ilia Ponomarenko
\thanks{Partially supported by RFFI, grant 02-01-00093}\\[-1pt]
\small Steklov Institute of Mathematics,\\[-3pt]
\small Fontanka 27, St. Petersburg 191011, Russia\\[-3pt]
{\tt \small inp@pdmi.ras.ru}\\[-3pt]
\small http://www.pdmi.ras.ru/\~{}inp
}
\date{28.02.2003}
\maketitle

\begin{abstract}
In this paper homomorphic cryptosystems are designed for
the first time over any finite group. Applying Barrington's
construction we produce for any boolean circuit of the logarithmic
depth its encrypted  simulation of a polynomial size over an
appropriate finitely generated  group.
\end{abstract}

\section{Homomorphic cryptography over groups}

\subs{\bf Definitions and results.}
An important problem of modern cryptography concerns secret public-key
computations in algebraic structures. There is a lot of public-key cryptosystems
using groups (see e.g. \cite{B,G,GP,KL,KMOV,MW,NS,Ra,Ri} and also
Subsection~\ref{sbs13}) but only a few of them have a homomorphic property
in the sense of the following definition (cf.~\cite{GP}).

\dfntnl{md}
Let $H$ be a finite nonidentity group, $G$ a finitely generated group
and $f:G\to H$ an epimorphism. Suppose that $R$ is a right transversal
of $\ker(f)$ in $G$, $A$ is a set  and $P:A\to G$ is a mapping such that
$\im(P)=\ker(f)$. A triple $\S=(A,P,R)$ is called a
{\it homomorphic cryptosystem} over $H$ with respect to $f$, if the following
conditions are satisfied for a certain integer $N\ge 1$ (called the size of
$\S$):
\nmrt
\tm{H1} the elements of the set $A$ are represented by words in a certain alphabet;
one can get randomly an element of $A$ of size $N$ within probabilistic time
$N^{O(1)}$,
\tm{H2} the elements of the group $G$ are represented by words in a certain alphabet;
one can test the equality of elements in $G$ and  perform group operations in $G$
(taking the inverse and computing the product) in time $N^{O(1)}$, provided that the sizes
of corresponding words are at most~$N$,
\tm{H3} the set $R$, the group $H$ and the bijection $R\to H$ induced
by $f$, are given by the list of elements, the multiplication table and the list
of pairs $(r,f(r))$, respectively;  $|R|=|H|=O(1)$,
\tm{H4} the mapping $P$ is a trapdoor function (cf.~\cite{GB}), i.e.
given a word $a\in A$ of the length $|a|$ an element  $P(a)$ can be computed
within probabilistic time $|a|^{O(1)}$, whereas the problem $\MM(P)$ is
computationaly hard,  while it can be solved by means of some additional secret
information,
\enmrt
\edfntn
where for any mapping $P:A\to G$ we define  $\MM(P)$ to be the problem of
testing whether given $g\in G$ belongs to $\im(P)$ and yielding a random element
$a\in A$  such that $P(a)=g$  whenever $g\in\im(P)$.

\rmrk
Having random generating in the set $A$ one can easily generate elements
of the group $G$ in a form $P(a)r$, $a\in A$, $r\in R$.
\ermrk

In a homomorphic cryptosystem $\S$ the elements of $H$ playing the role of the
alphabet of plaintext messages are publically encrypted in a probabilistic manner
by the elements of $G$ playing the role of the alphabet of ciphertext messages,
all the computations are performed in $G$ and the result is decrypted to $H$. More
precisely:

\vspace{4mm}
\noindent{\bf Public Key:} homomorphic cryptosystem $\S$.

\vspace{2mm}
\noindent{\bf Secret Key:} $\MM(P)$.

\vspace{2mm}
\noindent{\bf Encryption:} given a plaintext $h\in H$ encrypt as follows:
take $r\in R$ such that $f(r)=h$ (invoking (H3)) and a random element
$a\in A$ (using (H1); the ciphertext of $h$ is the element $P(a)r$ of~$G$
(computed by means of (H2) and (H4)).

\vspace{2mm}
\noindent{\bf Decryption:} given a cyphertext $g\in G$ decrypt as follows:
find the elements $r\in R$ and $a\in A$ such that $gr^{-1}=P(a)$ (using (H4));
the plaintext of $g$ is  the element $f(r)$ of~$H$ (computed by means of (H3)).
\vspace{4mm}

The main result of the present paper consists in the construction of a homomorphic
cryptosystem over arbitrary finite nonidentity group; the security of it is based on the
difficulty  of the following slight generalization of the factoring problem $\FA(n,m)$:
given a positive integer $n=pq$ with $p$ and $q$ being primes (of the same size),
a number $m\ge 2$ of a constant size such that $G_{n,m}/(\ZZ_n^*)^m\cong\ZZ_m^+$ where
$G_{n,m}=\{g\in \ZZ_n^*:\ {\bf J}_n(g)\in\{1,(-1)^{m\,(\mod 2)}\}\}$ with ${\bf J}_n$ being the
Jacobi symbol, and a transversal of $(\ZZ_n^*)^m$ in $G_{n,m}$, find the numbers $p,q$.
In addition, we assume that $m|p-1$ and $\GCD(m,q-1)=\GCD(m,2)$.
% although one could get rid of this extra assumption.

\thrml{mres}
Let $H$ be a finite nonidentity group  and $N\in\NN$.
Then one can design  a homomorphic cryptosystem $\S(H,N)$ of the size $O(N)$
over the group $H$;  the problem $\MM(P)$  where $P$ is  the trapdoor function,  is
probabilistic polynomial time equivalent to the problems  $\FA(n,m)$ for
appropriate $n=\exp(O(N))$ and $m$ running over the divisors of $|H|$.
\ethrm

First this result is proved for a cyclic group~$H$ (see Section~\ref{expc}), in this
case the group $G$ being a finite Abelian  group. Then in Section~\ref{mcn} a
homomorphic cryptosystem is yielded for an arbitrary~$H$, in this case the group
$G$ being a free product of certain Abelian groups produced in Section~\ref{expc}.
In Section~\ref{sesim} we recall the result from~\cite{Ba} designing  a polynomial
size simulation of any boolean circuit $B$ of the logarithmic depth over an arbitrary
unsolvable group $H$ (in particular, one can take $H$ to be the symmetric group
$\sym(5)$). Combining this result with Theorem~\ref{mres} provides an  {\it encrypted
simulation}  of $B$ over the group $G$: the output of this  simulation at a particular
input is a certain element $g\in G$, and thereby to know the output of $B$ one has
to be able to calculate $f(g)\in H$, which is supposedly to be difficult due to
Theorem~\ref{mres}. We mention that a different approach to encrypt boolean
circuits was undertaken in~\cite{SYY}.

\subsn\label{susectp}{\bf Discussion on complexity and security.}
One can see that the encryption procedure can be performed by means of public
keys efficiently. However, the decryption procedure is a secret one in the following
sense. To find the element $r$ one has to solve in fact, the membership problem
for the subgroup $\ker(f)$ of the group~$G$. We assume that a solution for each
instance $g'\in\ker(f)$ of this problem must have a ``proof'', which is actually an
element $a\in P^{-1}(g')$. Thus, the secrecy of the system is based on the assumption
that finding an element in the set $P^{-1}(g')$ i.e. solving $\MM(P)$ is an intractable
computation problem. On the other hand, our ability to compute $P^{-1}$ enables us to
efficiently implement the decryption algorithm. One can treat $P$ as a proof
system for membership to $\ker(f)$ in the sense of~\cite{CR}. Moreover, in
case when $A$ is a certain group and $P$ is a homomorphism we have the
following {\it exact} sequence of group homomorphisms
\qtnl{plkjh}
A\,\mapright{P}\,G\,\mapright{f}\,H\,\mapright{}\,\{1\}
\eqtn
(recall that the exact sequence means that the image of each homomorphism in
it coincides with the kernel of the next one).

The usual way in the public-key cryptography of providing an evidence of
the security of a cryptosystem is to fix a certain type of an attack (being an algorithm)
of cryptosystems and to prove that a cryptosystem is resistant with respect
to this type of an attack. The resistancy means usually that breaking a
cryptosystem with the help of the fixed type of an attack implies a certain
statement commonly believed to be unplausible. The most frequently used
in the cryptography  such statement (which we involve as well) is the
possibility to factorize an integer being a product of a pair of primes. Thus a
type of an attack we fix is that to break a homomorphic cryptosystem means to
be able to solve $\MM(P)$ (in other words, reveal the trapdoor).

Notice that in the present paper the group $H$ is always rather small, while the
group $G$ could be infinite but being always finitely generated. However, the
infinitness of~$G$ is not an obstacle for performing algorithms of encrypting
and decrypting (using the trapdoor information) since $G$ is a free product of
groups of a number-theoretic nature like $Z_n^*$; therefore one can easily verify
the condition (H2) and on the other hand this allows one to provide evidence for
the difficulty of  a decryption. In this connection we mention a public-key cryptosystem
from  \cite{DJSS} in which $f$ was the natural epimorphism from a free group $G$ onto
the group $H$  (infinite, non-abelian in general) given by generators
and relations. In this case for any element of $H$ one can produce its preimages
(encryptions) by inserting in a word (being already a produced preimage of $f$)
from $G$ any relation defining $H$. In other terms, decrypting of $f$ reduces to
the word problem in $H$. In our approach the word problem is solvable easily
due to a special presentation of the group $G$ (rather than given by generators
and relations).

\subsn\label{sbs13}{\bf Cryptosystems based on groups.}
To our best knowledge all known at present homomorphic cryptosystems
are more or less modifications of the following one. Let $n$ be the product
of two distinct large primes of size of the order $\log n$. Set
$G=\{g\in \ZZ_n^*:\ {\bf J}_n(g)=1\}$  and $H=\ZZ_2^+$. Then given a non-square
$r \in G$ the triple $(A,P,R)$ where
$$
R=\{1,r\},\quad A=\ZZ^*_n,\quad P(g):g\mapsto g^2,
$$
is a homomorphic cryptosystem over $H$ with respect to the natural epimorphism
$f:G\to H$ with $\ker(f)=\{g^2:\ g\in \ZZ_n^*\}$ (see~\cite{GM,GB}). We call it the
{\it quadratic residue cryptosystem}. It can be proved (see \cite{GM,GB})
that in this case solving the problem $\MM(P)$ is not easier than factoring $n$, whereas
given a prime divisor of $n$ this problem  can be solved  in probabilistic
polynomial time in $\log n$.

It is an essential assumption (being a shortcoming) in the quadratic residue
cryptosystem as well as other cryptosystems cited below that its security
relies on a fixed a priori (proof system) $P$. Indeed, it is not excluded
that an adversary could verify whether an element of $G$ belongs to $\ker (f)$
avoiding making use of~$P$, for example, in case of the quadratic residue
cryptosystem that would mean verifying that $g\in G$ is a square without
providing a square root of~$g$. Although, there is a common conjecture that
verifying for an element to be a square (as well as some power) is also
difficult.

Let us mention that a cryptosystem from \cite{P99} over $H=\ZZ_n^+$ (for the
same assumptions on $n$ as in the quadratic residue cryptosystem) with respect
to the homomorphism $f:G\to H$ where $G=\ZZ_{n^2}^*$ and
$\ker(f)=\{g^n:\ g\in G\}$, in which $A=G$ and $P:g\mapsto g^n$, is not
homomorphic in the sense of Definition~\ref{md} because condition (H3) of it
does not hold. (In particular, since $|G|\le |H|^2$, one can inverse $P$ in a polynomial
time in $|H|$.) By the same reason the cryptosystem
from \cite{OU98} over $H=\ZZ_{p}^+$ with respect to the homomorphism
$f:G\to H$ where $G=\ZZ_{p^2q}^*$ and $\ker(f)=\{g^{pq}:\ g\in G\}$
(here the integers $p,q$ are distinct large primes of the same size) is also
not homomorphic (besides, in this system only a part of the group $H$ is
encrypted). Some cryptosystems over certain dihedral groups were studied
in~\cite{Ra}. More general, in~\cite{GP} homomorphic cryptosystems were
designed over an arbitrary nonidentity solvable group.

We note in addition that an alternative setting of a homomorphic (in fact,
isomorphic) encryption $E$ (and a decryption $D=E^{-1}$) was proposed
in~\cite{KMOV}. Unlike Definition~\ref{md} the encryption $E:G\rightarrow G$
is executed in the same set $G$ (being an elliptic curve over the ring~$\ZZ_n$)
treated as the set of plaintext messages. If $n$ is composite, then $G$ is not
a group while being endowed with a partially defined binary operation which
converts $G$ in a group when $n$ is prime. The problem of decrypting this
cryptosystem is  close to the factoring of~$n$. In
this aspect \cite{KMOV} is similar to the well-known RSA scheme (see e.g.
\cite{GB}) if to interprete RSA as a homomorphism (in fact, isomorphism)
$E:Z_n^*\rightarrow Z_n^*$, for which the security relies on the difficulty
of finding the order of the group $Z_n^*$.

We complete the section by mentioning some cryptosystems using groups
but not being homomorphic in the sense of Definition~\ref{md}. The well-known
example is a cryptosystem which relies on the Diffie-Hellman key agreement
protocol (see e.g.~\cite{GB}).
It involves cyclic groups and relates to the discrete logarithm
problem~\cite{MW}; the complexity of this system was studied in~\cite{Sp}.
Some generalizations of this system to non-abelian groups (in particular,
the matrix groups over some rings) were suggested in~\cite{PKHK} where
secrecy was based on an analog of the discrete logarithm problems in groups
of inner automorphisms. Certain variations of the Diffie-Hellman systems over the
braid groups were described in~\cite{KL}; here several trapdoor one-way
functions connected with the conjugacy and the taking root problems in the
braid groups were proposed. Finally it should be noted that a cryptosystem
from~\cite{NS} is based on a monomorphism $\ZZ_m^+\to\ZZ_n^*$ by
means of which $x$ is encrypted by $g^x\,(\mod n)$ where $n,g$ constitute
a public key; its decrypting relates to the discrete logarithm problem and is
feasible in this situation due to a special choice of $n$ and $m$ (cf. also~\cite{B}).

\section{Homomorphic cryptosystems over cyclic groups}
\label{expc}

In this section we present an explicit homomorphic cryptosystem over a cyclic
group of an order $m>1$ whose decription is based on taking $m$-roots in the
group $\ZZ^*_n$
for a suitable $n\in\NN$. It can be considered in a sense as a generalization
of the quadratic residue cryptosystem over $\ZZ_2^+$.  Throughout
this section given $n\in\NN$ we denote by $|n|$ the size of the number~$n$.

Given a positive integer $m>1$ denote by $D_m$ the set of all pairs $(p,q)$ where
$p$ and $q$ are distinct odd primes such that
\qtnl{p0k0}
p-1=0\,(\mod m)\quad\textstyle{\rm and}\quad \GCD(m,q-1)=\GCD(m,2).
\eqtn
Let $(p,q)\in D_m$, $n=pq$ and $G_{n,m}$ be a  group defined by
\qtnl{p0k1}
G_{n,m}=\{g\in \ZZ_n^*:\ {\bf J}_n(g)\in\{1,(-1)^{m\,(\mod 2)}\}\}.
\eqtn
Thus $G_{n,m}=\ZZ_n^*$ for an odd $m$ and $[\ZZ^*_n:G_{n,m}]=2$ for an even $m$. In
any case this group  contains each  element $h=h_p\times h_q$ such that
$\lg h_p\rg=\ZZ_p^*$ and $\lg h_q\rg=\ZZ^*_q$  where $h_p$ and $h_q$  are the
$p$-component and the $q$-component of $h$ with respect to the
canonical decomposition $\ZZ_n^*=\ZZ_p^*\times\ZZ_q^*$. From (\ref{p0k0}) it follows that
$m$ divides the order of any such element $h$ and $\{1,h,\cdots,h^{m-1}\}$ is a transversal
of the group $G_{n,m}^m=\{g^m:\ g\in G_{n,m}\}$ in $G_{n,m}$. This implies that
$G_{n,m}/G_{n,m}^m\cong \ZZ_m^+$ where  the corresponding epimorphism is given by
the mapping
$$
f_{n,m}:G_{n,m}\to\ZZ_m^+,\quad g\mapsto i_g
$$
with $i_g$ being the element of $\ZZ_m^+$ such that $g\in G_{n,m}^mh^{i_g}$. From
(\ref{p0k0}) it follows that $\ker(f_{n,m})=G_{n,m}^m=\im(P_{n,m})$ where
$$
P_{n,m}:A_{n,m}\to G_{n,m},\quad g\mapsto g^m
$$
is a homomorphism from the group $A_{n,m}=\ZZ^*_n$ to the group $G_{n,m}$. In particular,
we have the exact sequence (\ref{plkjh}) with $A=A_{n,m}$, $P=P_{n,m}$, $f=f_{n,m}$,
$G=G_{n,m}$ and $H=\ZZ_m^+$.  Next, it is easily seen that  any element of the set
$$
\R_{n,m}=\{R\subset G_{n,m}:\ |f_{n,m}(R)|=|R|=m\}
$$
is a right transversal of $G_{n,m}^m$ in $G_{n,m}$. We notice that
by the Dirichlet theorem on primes in arithmetic progressions
(see e.g.~\cite{D}) the set $D_m$ is not empty. Moreover, by the same reason
the set
$$
D_{N,m}=\{n\in\NN:\ n=pq,\ (p,q)\in D_m,\ |p|=|q|=N\}
$$
is also nonempty for sufficiently large $N\in\NN$.

\thrml{ehcc}
Let $H$ be a cyclic group of order $m>1$. Then given $N\in\NN$ and $n\in D_{N,m}$
one can design  a homomorphic cryptosystem $\S_n(H,N)$ of the size $O(N)$ over the
group $H$;  the problem $\MM(P)$ where $P$ is  the trapdoor function,  is probabilistic
polynomial time equivalent to the problem $\FA(n,m)$.
\ethrm
\proof First we desribe a probabilistic polynomial time algorithm which yields a certain
$n\in D_{N,m}$. The algorithm picks randomly integers $p=1\,(\mod m)$ and
$q=-1\,(\mod m)$ from the interval $[2^N,2^{N+1}]$ and tests primality of the picked
numbers by means of e.g.~\cite{SS}. According to~\cite{D} there is a constant $c>0$
such that for any $b$ relatively prime
with~$m$ there are at least $c2^N/(\varphi(m)N)$ primes of the form $mx+b$ in the
interval $[2^N,2^{N+1}]$. Therefore, after $O(N)$ attempts the algorithm would yield  a
pair $(p,q)\in D_m$ with a probability greater than $2/3$ (actually, one can replace $2/3$
by an arbitrary constant less than~$1$). Thus given $N\in\NN$ one can design in
probabilistic time $N^{O(1)}$ a number $n\in D_{N,m}$, a random element $R\in\R_{n,m}$
(see e.g.~\cite{NS}) and the triple
\qtnl{hcfcg}
\S_n(H,N)=(A,P,R)
\eqtn
where $A=A_{n,m}$ and $P=P_{n,m}$ (below without loss of generality we assume that
$H=\ZZ_m^+$).

We will show that for any  $n\in D_{N,m}$ and  $R\in\R_{n,m}$ the triple
$\S_n(H,N)$ is a homomorphic cryptosystem of the size $O(N)$
over the group $H$ with respect to the epimorphism $f:G\to H$ where  $f=f_{n,m}$ and
$G=G_{n,m}$. For this purpose we note that in this case there is  the exact sequence (\ref{plkjh})
(see above). Next, we will  represent the elements of the set $A$ and of the group $G$ by
integers modulo $n$, and those of the group $H$ by integers modulo $m$. Then  conditions
(H1), (H2) and (H3) of Definition~\ref{md} are trivially satisfied. Since the epimorphism
$P$ is obviously a polynomial time computable one, it suffices to verify condition (H4), i.e.
that the problems $\MM(P)$ and $\FA(n,m)$ are probabilistic  polynomial time equivalent.

Suppose that  we are given an algorithm solving the problem $\FA(n,m)$. Then we can find
the decomposition $n=pq$. Now using  Rabin's probabilistic polynomial-time algorithm for
finding roots of polynomials over finite prime fields (see~\cite{R}), we can solve the problem
$\MM(P)$ for an element $g\in G$  as follows:

\begin{itemize}
\item[]{\bf Step 1.} Find the numbers $g_p\in\ZZ_p^*$ and $g_q\in\ZZ_q^*$
such that $g=g_p\times g_q$, i.e. $g_p=g\ \,(\mod p)$, $g_q=g\ \,(\mod q)$.
\vspace{2mm}

\item[]{\bf Step 2.} Apply Rabin's algorithm for the field of order $p$ to
the polynomial $x^m-g_p$ and for the field of order $q$ to the polynomial
$x^m-g_q$. If at least one of this polynomials has no roots, then output
``$P^{-1}(g)=\emptyset$''; otherwise let $h_p$ and $h_q$ be corresponding
roots.
\vspace{2mm}

\item[]{\bf Step 3.}
Output ``$P^{-1}(g)\ne\emptyset$'' and $h=h_p\times h_q$.
\end{itemize}
\vspace{2mm}

\noindent We observe that the set $P^{-1}(g)$ is empty, i.e. the $g$ is not an $m$-power
in $G$, iff  at least one of the elements $g_p$ and $g_q$ found at Step 1 is not an $m$-power
in $\ZZ^*_p$ and $\ZZ^*_q$ respectively. This implies the correctness of the output at Step~2.
On the other hand, if the procedure terminates at Step~3, then
$h^m=h_p^m\times h_q^m=g_p\times g_q=g$, i.e. $h\in P^{-1}(g)$. Thus
the problem $\MM(P)$ is reduced to the problem $\FA(n,m)$ in probabilistic
time $N^{O(1)}$.

Conversely, suppose that  we are given an algorithm solving the problem $\MM(P)$. Then the
following procedure using well-known observations \cite{GB} enables us to find the decomposition
$n=pq$.

\begin{itemize}
\item[]{\bf Step 1.} Randomly choose $g\in\ZZ_n^*$. Set $T=\{g\}$.
\vspace{2mm}

\item[]{\bf Step 2.}  While $|T|<3-(m\,(\mod 2))$, add to $T$ a random $m$-root of the
element $g^m$ yielded  by  the algorithm for the problem $\MM(P)$.
\vspace{2mm}

\item[]{\bf Step 3.}  Choose $h_1,h_2\in T$ such that $q=\GCD(h_1-h_2,n)\ne 1$.
Output $q$ and $p=n/q$.
\end{itemize}
\vspace{2mm}

\noindent To prove the correctness of the procedure  we observe that there exists at
least~$2$ (resp.~$4$) different $m$-roots of the element $g^m$ for odd $m$ (resp. for even
$m$) where $g$ is the element chosen at Step~1. So the loop at Step 2 and hence the entire
procedure  terminates with a large probability after a polynomial number of
iterations. Moreover, let $T_q=\{h_q:\ h\in T\}$ where $h_q$ is the $q$-component of $h$.
Then from (\ref{p0k0}) it follows that $|T_q|=1$ for odd $m$, and $|T_q|\le 2$ for even $m$.
Due to the construction of $T$ at Step~2 this implies that there exist different elements
$h_1,h_2\in T$  such that $(h_1)_q=(h_2)_q$, and consequently
$$
h_1=(h_1)_q=(h_2)_q=h_2\ \,(\mod q).
$$
Since  $h_1\ne h_2\ \,(\mod n)$, we conclude that $h_1-h_2$ is a multiple of $q$ and output at
Step~3 is correct.\bull

We complete the section by mentioning that the decryption algorithm of the
homomorphic cryptosystem $\S_{N,m,n}$ can be slightly modified
to avoid  applying Rabin's algorithm for finding roots of polynomials over finite
fields. Indeed, it is easy to see that an element $g=g_p\times g_q$ of the group
$G$ belongs to the group $G^m$  iff $g_p^{(p-1)/m}=1\,(\mod p)$ and
$g_q^{(q-1)/m'}=1\,(\mod q)$ where $m'=\GCD(m,q-1)$.

\section{Homomorphic cryptosystems using free products}
\label{mcn}
Throughout the section we denote by $W_X$ the set of all the
words $w$ in the alphabet~$X$; the length of $w$ is denoted
by $|w|$. We use the notation $G=\lg X;\R\rg$ for a presentation of a group
$G$ by the set $X$ of generators and the set $\R$ of relations. Sometimes
we omit $\R$ to stress that the group $G$ is generated by the set~$X$.
The unity of $G$ is denoted by $1_G$ and we set $G^\#=G\setminus\{1_G\}$.
Finally,  given a positive integer $n$ we set $\ov n=\{1,\ldots,n\}$.

\subs\label{sss1}{\bf Calculations in free products of groups.}
Let us remind the basic facts on free products of groups (see e.g. \cite[Ch.~4]{MKS}).
Let $G_1,\ldots,G_n$ be finite groups, $n\ge 1$. Given a presentation
$G_i=\lg X_i;\R_i\rg$, $i\in\ov n$, one can form a group
$G=\lg X;\ \R\rg$ where $X=\cup_{i\in\ov n} X_i$ (the disjoint union) and
$\R=\cup_{i\in\ov n}\R_i$. It can be proved that  this group does not
depend on the choice of presentations of~$\lg X_i;\R_i\rg$, $i\in\ov n$. It is
called the {\it free product} of the groups $G_i$ and is denoted by
$G=G_1*\cdots*G_n$; one can see that it does not depend on the order of
factors. Without loss of generality we assume below that
$G_i$ is a subgroup of $G$ and $X_i=G_i^\#$ for all~$i$.
In this case $G\subset W_X$ and $1_G$ equals the empty word of $W_X$.
Moreover, it can be proved that
\qtnl{l257}
G=\{x_1\cdots x_k\in W_X:\ x_j\in G_{i_j}\ \textstyle{\rm for}\ j\in\ov k,\
\textstyle{\rm and}\
i_j\ne i_{j+1}\ \textstyle{\rm for}\
j\in\ov{k-1}\}.
\eqtn
Thus each element of $G$ is a word of $W_X$ in which no two
adjacent letters belong to the same set among the sets~$X_i$, and any
two such different  words are different elements of~$G$. To
describe the multiplication in $G$ let us first define recursively the
mapping $W_X\to G$, $w\mapsto\ov w$ as follows
\qtnl{losh}
\ov w=\cases{w, &if $w\in G$,\cr
\ov{\ldots(x\cdot y)\ldots}, &if $w=\ldots xy\ldots$ with $x,y\in X_i$
for some $i\in\ov n$,\cr}
\eqtn
where $x\cdot y$ is the product of $x$ by $y$ in the group $G_i$. One can
prove that the word $\ov w$ is uniquely determined by~$w$ and so the
mapping is correctly defined. In particular, this implies that given $i\in\ov n$
we have
\qtnl{los1}
\ov{x_1\cdots x_k}\in G_i\ \Leftrightarrow\ \ov{x_1\cdots x_k}=\ov{x_{j_1}\cdots x_{j_{k'}}}
\eqtn
where $\{j_1,\ldots,j_{k'}\}=\{j\in\ov k:\ x_j\in G_i\}$. Now given $g,h\in G$ the product of $g$
by $h$ in $G$ equals $\ov{gh}$.

\lmml{lmn1}
Let $G=G_1*\cdots *G_n$, $K=K_1*\cdots *K_n$ be groups and  $f_i$ be an
epimorphism from $G_i$ onto $K_i$, $i\in\ov n$.  Then the mapping
\qtnl{l1}
\varphi:G\to K,\quad
x_1\cdots x_k\mapsto \ov{f_{i_1}(x_1)\cdots f_{i_k}(x_k)}
\eqtn
where $x_j\in G_{i_j}$,  $j\in\ov k$, is an epimorphism. Moreover,
$\varphi|_{G_i}=f_i$ for all $i\in\ov n$.
\elmm
\proof Since $K=\lg Y\rg$ where $Y=\cup_{i\in\ov n}K_i^\#$, the
surjectivity of the mapping $\varphi$ follows from the surjectivity of the
mappings~$f_i$, $i\in\ov n$.
Next, let $\varphi_0:W_X\to W_Y$ be the mapping taking
$x_1\cdots x_k$  to $f_{i_1}(x_1)\cdots f_{i_k}(x_k)$.
Then  it is easy to see that $\varphi(g)=\ov{\varphi_0(g)}$ for all  $g\in G$ and
$\varphi_0(ww')=\varphi_0(w)\varphi_0(w')$ for all $w,w'\in W_X$.
Since $\ov{\ov w\,\ov w'}=\ov{ww'}$ for all $w,w'\in W_X$, this implies that
$$
\ov{\varphi(g)\varphi(h)}=
\ov{\ov{\varphi_0(g)}\,\ov{\varphi_0(h)}}=
\ov{\varphi_0(g)\varphi_0(h)}=
\ov{\varphi_0(gh)}=
\varphi(\ov{gh})
$$
for all $g,h\in G$. Thus the mapping $\varphi$ is a homomorphism.
Since obviously $\varphi|_{G_i}=f_i$ for all $i\in\ov n$, we are done.\bull

Let $H$ be a finite nonidentity group and $K$ be the free product of cyclic groups
generated by all the nonidentity elements of $H$. Set
$$
\R^{(0)}=\{h^{(m_h)}\in W_{H^\#}: h\in H^\#\},
$$
$$
\R^{(1)}=\{h^{(i)}h'\in W_{H^\#}:\ h,h'\in H^\#,\ 0<i<m_h,\ h^i\cdot h'=1_H\},
$$
$$
\R^{(2)}=\{hh'h''\in W_{H^\#}:\ h,h',h''\in H^\#,\ h'\not\in\lg h\rg,\
h\cdot h'\cdot h''=1_H\}
$$
where $h^{(i)}$ is the word of length $i\ge 1$ with all letters being equal~$h$,
$m_h$ is the order of $h\in H$ and $\cdot$ denotes the multiplication in $H$. Then
one can see that
\qtnl{presk}
K=\lg H^\#;\R^{(0)}\rg
\eqtn
and there is the natural epimorphism $\psi':K\to H'$ where
$H'=\lg H^\#;\R^{(0)}\cup\R^{(1)}\cup\R^{(2)}\rg$. Since relations belonging to
$\R^{(i)}$, $i=0,1,2$, are satisfied in $H$, we conclude that
$\ker(\psi')h_1\ne\ker(\psi')h_2$ whenever $h_1$ and $h_2$ are
different elements of~$H$ (we identify $1_K$ and $1_H$). On the other hand,
it is easy to see that any right coset of $K$ by $\ker(\psi')$ contains a word
of length at most~1, i.e. an element of~$H$. Thus $K=\cup_{h\in H}\ker(\psi')h$,
the mapping
\qtnl{o123o}
\psi:K\to H,\quad k\mapsto h_k
\eqtn
where $h_k$ is the uniquely detemined element of $H$ for which
$k\in\ker(\psi')h_k$, is an epimorphism and $\ker(\psi)=\ker(\psi')$.

\subsn\label{mcoac}{\bf Main construction of a homomorphic cryptosystem.}
Let $H$ be a finite nonidentity group and $N$ be a positive integer. We
are going to describe a homomorphic cryptosystem $\S(H,N)$ of size
$O(N)$ over the group~$H$. Suppose first that
$H$ is a cyclic group of an order $m>1$. Then we set $\S(H,N)=\S_n(H,N)$
where $n\in D_{N,m}$ (see Theorem~\ref{ehcc}).  If $H$ is not a cyclic group,
then $\S(H,N)$ is defined as follows.

Let $H^\#=\{h_1,\ldots,h_n\}$ where $n$ is a positive integer (clearly,
$n\ge 3$). Set $D_{N,H}=\cup_{i\in\ov n}D_{N,m_i}$ where $m_i$ is the order
of the group $K_i=\lg h_i\rg$. Given $i\in\ov n$ choose $n_i\in D_{N,m_i}$
and set $\S_i=(A_i,P_i,R_i)$ to be the homomorphic cryptosystem $\S_{n_i}(K_i,N)$
with respect to the epimorphism $f_i:G_i\to K_i$ (see  Theorem~\ref{ehcc}).  Without
loss of generality we assume that $G_i$ is a subgroup of the group $\ZZ^*_{n_i}$.
Set
\qtnl{abc1}
G=G_1*\cdots*G_n,\quad f=\psi\circ\varphi,
\eqtn
where the mappings $\varphi$ and $\psi$ are defined by (\ref{l1}) and (\ref{o123o})
respectively, with $K=K_1*\cdots*K_n$. From Lemma~\ref{lmn1} and the definition
of~$\psi$ it follows that the mapping  $f:G\to H$  is an epimorphism from $G$
onto~$H$.

To define a proof system for membership to $\ker(f)$ (see
Subsection~\ref{susectp}) we  set
\qtnl{eoin}
X_\varphi=X\cup A_0\quad
X=\cup_{i\in\ov n}G_i\setminus\ker(f_i),\quad
A_0=\cup_{i\in\ov n}A_i,
\eqtn
all the unions are assumed to be the disjoint ones. Denote by $\rightarrow$ the transitive
closure of the binary relation $\Rightarrow$ on  the set $W_{X_\varphi}$ defined by
\qtnl{spl2}
v\Rightarrow  w\quad\textstyle{\rm iff}\quad w=x^{-1}x_0vx,\qquad v,w\in W_{X_\varphi}
\eqtn
where $x\in X\cup\{1_A\}$ and $x_0\in A_0\cup\{1_A\}$ with $1_A$ being  the empty word
of $W_{X_\varphi}$. Thus $v\rightarrow w$ if there exist words $w_1=v,w_2,\ldots,w_l=w$
of $W_{X_\varphi}$ such that $w_i\Rightarrow w_{i+1}$ for $i\in\ov{l-1}$.
We set
\qtnl{ool1}
A_\varphi=\{a\in W_{X_\varphi}:\ 1_{A_\varphi}\rightarrow a\},\quad
P_\varphi:A_{\varphi}\to G,\
a_1\cdots a_k\mapsto \ov{P_\varphi(a_1)\cdots P_\varphi(a_k)}
\eqtn
where $P_\varphi|_X=\id_X$ and $P_\varphi|_{A_i}=P_i$ for all~$i$. We observe that if
$\ov v\in\ker(\varphi)$ and $v\Rightarrow w$ for some $v,w\in W_{X_\varphi}$ then obviously
$\ov w\in\ker(\varphi)$ (see~(\ref{spl2})). By induction on the size of a word this implies
that $P_\varphi(A_\varphi)\subset\ker(\varphi)$. Next,  set
\qtnl{ool2}
A_\psi=\{r\in W_{R_\psi}:\ f(\ov r)=1_H\},\quad
P_\psi:A_\psi\to G,\ a\mapsto\ov a
\eqtn
where $R_\psi=\cup_{i\in\ov n}R_i$. It is easily seen that the restriction of $\varphi$ to the
set $R_\varphi=G\cap W_R$ induces a bijection from this set to the group $K$. This
shows that $R_\varphi$ is a right transversal of $\ker(\varphi)$ in $G$. Finally we define
\qtnl{ool3}
A=A_\varphi\times A_\psi,\quad
P:A\to G,\ (a,b)\mapsto\ov{P_\varphi(a)P_\psi(b)}.
\eqtn
Let $R$ be a right transversal of $\ker(f)$ in $G$, for instance one can take
$R=\{1_G\}\cup \{r'_i\}_{i\in\ov n}$ where $r'_i$ is the element of $R_i$
such that $\psi(r'_i)=h_i$, $i\in\ov n$. Set $\S(H,N)=(A,P,R)$.

\subsn{\bf Proof of Theorem~\ref{mres}.}

First we observe that if $H$ is a cyclic group, then the required statement follows
from Theorem~\ref{ehcc}. Suppose from now on that the group $H$ is not cyclic. Let
us describe the presentations of the set $A$ and the groups $G$ and $K$. Given
$i\in\ov n$ the  elements $a\in A_i$ and $g\in G_i$  being the elements of $\ZZ_{n_i}^*$
will be represented by the ``letters'' $]a,i[$ and $[g,i]$ respectively. This completely
defines the representations of the set $A$ and the group $G$.  We note that relying
on (\ref{spl2}), (\ref{ool1}) and (\ref{ool2}) one can randomly generate elements of~$A$.

The group $G$ is represented by the subset~(\ref{l257}) of the set $W_X$.
To multiply two elements $g,h\in G$ one has to find the word $\ov{gh}$ of $W_X$.
It is easy to see that this can be done by means of the recursive procedure (\ref{losh})
in time $((|g|+|h|)N)^{O(1)}$ (here $[x,i]\cdot [y,i]=[xy,i]$ for all $x,y\in\ZZ^*_{n_i}$ where
$xy$ is  the product modulo $n_i$ of the numbers $x$ and $y$, and  $n_i\le \exp^{O(N)}$
because $n_i\in D_{N,m_i}$). Since taking the inverse of $g\in G$ can be easily
implemented in time $(|g|N)^{O(1)}$, we will estimate further the running time of the
algorithms via the number of performed group operations in $G$ and via the sizes of
the involved operands.

Finally the group $H$ as well as the groups $K_i$, $i\in\ov n$, are given by their
multiplication tables, and the group $K$ is given by the presentation~(\ref{presk}).
Thus all the group operations in $K$ can be performed in  time polynomial in the lengths
of the input words belonging to~$W_{H^\#}$.

Now, we have the following sequence of the mappings:
$$
A_\varphi\times A_\psi \stackrel{P}{\longrightarrow}
G_1*\cdots *G_n\stackrel{\varphi}{\longrightarrow}
K_1*\cdots *K_n\stackrel{\psi}{\longrightarrow} H.
$$
In the following two lemmas we study the homomorphisms $\varphi$ and $\psi$ from
the algorithmic point of view.

\lmml{spl3}
For the mapping $P_\varphi$ defined in~(\ref{ool1}) the following statements hold:
\nmrt
\tm{i1} given $a\in A_\varphi$ the element $P_\varphi(a)$ can be found in time $|a|^{O(1)}$,
\tm{i2} $\im(P_\varphi)=\ker(\varphi)$,
\tm{i3} given an oracle  $Q_i$ for the problem $\MM(P_i)$ for all $i\in\ov n$,
the problem $\MM(P_\varphi)$ for $g\in G$ can be solved  by
means of at most $|g|^2$ calls of oracles $Q_i$, $i\in\ov n$,
\tm{i4} for each $i\in\ov n$ the problem $\MM(P_i)$ is polynomial time
reducible to the problem $\MM(P_\varphi)$.
\enmrt
\elmm
\proof Let us prove statement (i1). Let $a=a_1\cdots a_k$ be an element of $A_\varphi$.
To find $P_\varphi(a)$ according to (\ref{ool1}) we need to compute the words
$P_\varphi(a_j)$, $j\in\ov k$, and then to compute the word $\ov w$ where
$w=P_\varphi(a_1)\cdots P_\varphi(a_k)$. The first stage can be done in time $|a|^{O(1)}$
because each mapping $P_i$, $i\in\ov n$, is polynomial time computable due
to Section~\ref{expc}. Since the size of $w$ equals $|a|$,  the element
$P_\varphi(a)$ can be found within the similar time bound (one should take into account
that in the recursive procedure (\ref{losh}) applied for computing $\ov w$ from $w$ the
length of a current word decreases at each step of the procedure).

To prove statements~(i2) and (i3) we note first that the inclusion
$\im(P_\varphi)\subset \ker(\varphi)$ was proved after the definition of $A_\varphi$ and
$P_\varphi$ in~(\ref{ool1}). The converse inclusion as
well as statement (i3) will be proved by means of the following recursive procedure which
for a given element $g=x_1\cdots x_k$ of $G$ with $x_j\in G_{i_j}$ for $j\in\ov k$,
produces a certain pair $(a_g,t_g)\in A_\varphi\times G$. Below we show that this
procedure actually solves the problem $\MM(P_\varphi)$.

\begin{itemize}
\item[]{\bf Step 1.} If $g=1_G$, then output $(1_{A_\varphi},1_G)$.
\vspace{2mm}

\item[]{\bf Step 2.} If the set $J=\{j\in\ov k:\ x_j\in\ker(f_{i_j})\}$ is empty, then output
$(1_{A_\varphi},g)$.
\vspace{2mm}

\item[]{\bf Step 3.} Set $h=\ov{x_{j+1}\cdots x_kx_1\cdots x_{j-1}}$ where $j$ is the smallest
element of the set~$J$.
\vspace{2mm}

\item[]{\bf Step 4.} Recursively find the pair $(a_h,t_h)$. If $t_h\ne 1_G$,
then output $(a_h,t_h)$.
\vspace{2mm}

\item[]{\bf Step 5.} If $t_h=1_G$, then output $(a_g,1_G)$ where
$a_g=x_1\cdots x_{j-1}a_ja_hx_{j-1}^{-1}\cdots x_1^{-1}$ with $a_j$ being an arbitrary
element of $A_{i_j}$ such that $P_{i_j}(a_j)=x_j$.\bull
\end{itemize}
\vspace{2mm}

Since each recursive call at Step~4 is applied to the word $h\in G$ of size at most $|g|-1$,
the number of  recursive calls is at most $|g|$. So the total number of oracle $Q_i$
calls, $i\in \ov n$, at Step~2 does not exceed $|g|^2$. Thus the running time of the algorithm is
$(|g|)^{O(1)}$ and statements (i2), (i3) are consequences of the following lemma.

\lmml{l878}
$g\in\ker(\varphi)$ iff $t_g=1_G$. Moreover, if $t_g=1_G$, then
$a_g\in A_{\varphi}$ and $P_\varphi(a_g)=g$.
\elmm

\proof  We will prove the both  statements by induction on $k=|g|$. If $k=0$, then the
procedure terminates at Step~1 and we are done. Suppose that $k>0$.
If the procedure terminates at Step~2, then $t_g\ne 1_G$. In this case we have
$|\varphi(g)|=|g|=k>0$, whence $g\not\in\ker(\varphi)$. Let the procedure terminate
at Step~4 or at Step~5. Then $|h|\le|g|-1$ (see Step~3). So by the induction hypothesis
we can assume that $h\in\ker(\varphi)$ iff $t_h=1_G$. On the other hand, taking into account
that $x_j\in\ker(f_{i_j})$ (see the definition of $j$ at Step~3) we get  that
$h\in\ker(\varphi)$ iff  $\ov{u x_jhu^{-1}}\in\ker(\varphi)$  where $u=x_1\ldots,x_{j-1}$.
Since
\qtnl{kjd}
\ov{u x_jhu^{-1}}=
\ov{x_1\cdots x_{j-1}x_jhx_{j-1}^{-1}\cdots x_1^{-1}}=\ov{x_1\cdots x_k}=\ov g=g,
\eqtn
this means that $g\in\ker(\varphi)$ iff $h\in\ker(\varphi)$ iff $t_h=1_G$. This proves the first statement of
the lemma because $t_h=t_g$ due to Steps~4 and~5.

To prove the second statement, suppose that $t_g=1_G$.
Then the above argument shows that $h\in\ker(\varphi)$ and so $a_h\in A_\varphi$ and
$P_\varphi(a_h)=h$ by the induction hypothesis. This implies that
$1_{A_\varphi}\rightarrow a_h$. On the other
hand, from the definition of $a_g$ at Step~5 it follows that $a_h\rightarrow a_g$
(see~(\ref{spl2})). Thus $1_{A_\varphi}\rightarrow a_g$, i.e. $a_g\in A_\varphi$
(see~(\ref{ool1})). Besides, from the minimality of $j$ it follows that $x_l\in X$
(see (\ref{eoin})) and hence $P_\varphi(x_l)=x_l$ and $P_\varphi(x_l^{-1})=x_l^{-1}$
for all $l\in\ov{j-1}$ (see (\ref{ool1})). Since $P_\varphi(a_j)=x_j$ and
$\ov h=h=\ov{x_{j+1}\cdots x_kx_1\cdots x_{j-1}}$ (see Step~3), we obtain
by (\ref{kjd}) that
$$
P_\varphi(a_g)=\ov{u x_jP_\varphi(a_h)u^{-1}}=\ov{ux_jhu^{-1}}=g
$$
which completes the proof of the Lemma~\ref{l878}.\bull
\vspace{1mm}

To prove statement~(i4) let $i\in\ov n$ and $g\in G_i$. Then since obviously $g\in\ker(f_i)$
iff $g\in\ker(\varphi)$, one can test whether $g\in\ker(f_i)$  by means of an algorithm
solving the problem $\MM(P_\varphi)$. Moreover, if $g\in\ker(f_i)$, then this algorithm
yields an element $a\in A_\varphi$ such that $P_\varphi(a)=g$. Then assuming
$a=a_1\cdots a_k$ with $a_j\in X_\varphi$, the set $J_a=\{j\in\ov k:\ a_j=]a^*_j,i[\}$ can be
found in time $O(|a|)$ (we recall that due to our presentation any element  $a_j$ is
of the form either $]a_j^*,i_j[$ or $[a^*_j,i_j]$ where $i_j\in\ov n$ and $a_j^*\in\ZZ^*_{n_{i_j}}$, and
$P_{i_j}(a_j)\in\ker(f_{i_j})$ iff $a_j\in A_0$ iff $a_j=]a_j^*,i_j[$). Now the element
$$
a^*=]\prod_{j\in J_a}a^*_j,i[
$$
obviously belongs to the set $A_i\subset A_0$. On the other hand, since $g\in G_i$,
we get by (\ref{los1}) that
\qtnl{pmio}
g=\ov{P_\varphi(a_1)\cdots P_\varphi(a_k)}=\ov{\prod_{j\in J}P_\varphi(a_j)}
\eqtn
where $J=\{j\in\ov k:\ P_\varphi(a_j)\in G_i\}$. Taking into account that $G_i$ is an
Abelian group and the mapping $P_i:A_i\to G_i$ is a homomorphism, we have
\qtnl{pmio1}
\ov{\prod_{j\in J}P_\varphi(a_j)}=
\ov{\prod_{j\in J_a}P_i(a_j)\prod_{j\in J\setminus J_a}P_\varphi(a_j)}=
\ov{P_i(a^*)\prod_{j\in J\setminus J_a}P_\varphi(a_j)}.
\eqtn
Moreover, since $1_{A_\varphi}\rightarrow a$, from (\ref{spl2}) it follows that there exists involution
$j\to j'$ on the set $J\setminus J_a$ such that $a_j=[a_j^*,i]$ iff $a_{j'}=[(a_j^*)^{-1},i]$
(we recall that $a_j=]a^*_j,i[$ for $j\in J_a$ and $a_j=[a^*_j,i]$ for $j\in J\setminus J_a$).
This implies that $\prod_{j\in J\setminus J_a}P_\varphi(a_j)=1_G$. Thus from
(\ref{pmio}) and  (\ref{pmio1}) we conclude that:
$$
g=\ov{P_i(a^*)}=\ov{P_\varphi(a^*)}=P_\varphi(a^*).
$$
This shows that the element $a^* \in A_i$ with $P_\varphi(a^*)=g$ can be constructed
from $a$ in time $O(|a|)$. Using condition (H1) for the cryptosystem $\S_i$, one can
efficiently transform the element $a^*$ to a random  element $\widetilde a$ so that
$P_\varphi(\widetilde a)=P_\varphi(a^*)=g$. Thus  the problem $\MM(P_i)$  is polynomial
time reducible to the problem $\MM(P_\varphi)$. The Lemma~\ref{spl3} is
proved.\bull

\lmml{ortd}
Let $K$ be the group given by presentation~(\ref{presk}) and the epimorphism $\psi$
is defined by~(\ref{o123o}). Then given $k\in K$ one can find the element $\psi(k)$
in time $(|k| |H|)^{O(1)}$.
\elmm
\proof It is easy to see that the group $K$ can be identified with the subset of the
set $W_{H^\#}$ so that $w\in K$ iff  the length of any subword of $w$ of the form
$h\cdots h$ (i.e. the repetition of a letter $h$) is at most $m_h-1$.  Having this in mind
we claim that the following recursive procedure computes $\psi(k)$ for all
$k=x_1\cdots x_t\in K$.

\begin{itemize}
\item[]{\bf Step 1.} If $t\le 1$, then  output $\psi(k)=k$.
\vspace{2mm}

\item[]{\bf Step 2.} Choose $h\in H$ such that $x_1x_2h\in\R^{(1)}\cup\R^{(2)}$.
\vspace{2mm}

\item[]{\bf Step 3.} Output $\psi(k)=\psi(h^{-1}x_3\cdots x_t)$.
\end{itemize}
\vspace{2mm}

The correctness of the procedure follows from the definitions of sets $\R^{(1)}$, $\R^{(2)}$,
and the fact that recursion at Step~3 is always applied to a word the length of
which is smaller than the length of the current word. In fact, the above procedure
produces the representation of $k$ in the form $k=w_1\cdots w_{t-1}\psi(k)$ where
$w_j\in\R^{(1)} \cup\R^{(2)} $ for all $j\in\ov{t-1}$ and $\psi(k)\in H$. Since obviously
$w_1\cdots w_{t-1}\in\ker(\psi)$, we conclude that $\psi(k)=h_k$ (see~(\ref{o123o})).
To complete the proof it suffices to note that the running time of the above procedure
is $O(|k|(|\R^{(1)}|+|\R^{(2)}|))$.\bull

Finally, let us complete the proof of Theorem~\ref{mres}. First, we observe that by
Lemma~\ref{lmn1} the mapping $f:G\to H$ is a composition of two epimorphisms
and so is an  epimorphism too. Next, to prove that  the mapping $P:A\to\ker(f)$ is a
surjection, we recall that the set $R_\varphi$ defined after (\ref{ool2})
is a right  transversal of $\ker(\varphi)$ in $G$. So given $g\in\ker(f)$ there exist
uniquely determined elements $g_\varphi\in\ker(\varphi)$ and $r_\varphi\in R_\varphi$
such that $g=\ov{g_\varphi r_\varphi}$. Since
$$
1_H=f(g)=\psi(\varphi(\ov{g_\varphi r_\varphi}))=\psi(\varphi(r_\varphi))=f(r_\varphi),
$$
we see that $r_\varphi\in A_\psi$ (see~\ref{ool2}).  Besides, from statement (i2) of
Lemma~\ref{spl3} it follows that there exists $a\in A_\varphi$ for which
$P_\varphi(a)=g_\varphi$. Therefore, due to~(\ref{ool3}) we have
$$
P(a,r_\varphi)=\ov{P_\varphi(a)P_\psi(r_\varphi)}=\ov{g_\varphi r_\varphi}=g.
$$
Thus the mapping $P$ is a surjection. Since conditions (H1)-(H3) of the Definition~\ref{md}
are satisfied (see the end of Subsection~\ref{mcoac}),  it remains to verify the condition (H4),
i. e. that $P$ is a trapdoor function.

First, we observe that by statement~(i1) of Lemma~\ref{spl3} and by Lemma~\ref{ortd}
the mappings $P_\varphi$ and $P_\psi$ are polynomial time computable, whence so
does the mapping~$P$. Next, given an element $g\in G$ there exists the uniquely
determined element $r\in R$ such that $f(g)=f(r)$ or, equivalently,  $f(gr^{-1})=1_H$.
Since $|R|=O(1)$, this implies that the problem of the computation of the epimorphism
$f$ is polynomial time equivalent to the problem of recognizing elements of $\ker(f)$
in $G$, i. e. in our setting to the problem $\MM(P)$. Thus, we have to
show that
\nmrt
\tm{a} the problem $\MM(P)$ can be efficiently solved by means of using the  trapdoor
information for the homomorphic cryptosystems $(R_i,A_i,P_i)$,  $i\in\ov n$, i.e. the
factoring of integers $n_i\in D_{n,m_i}$,
\tm{b} for any $i\in\ov n$ the problem $\MM(P_i)$  (to which the factoring of integers
$n_i$ is reduced) is polynomial time reducible to the
problem $\MM(P)$.
\enmrt
Suppose that for each $i\in\ov n$ there is an oracle  for  the problem
$\MM(P_i)$. Then  given $g_i\in G_i$ one can find the element
$f_i(g_i)$ in time $N^{O(1)}$. So given $g\in G$ the element $k=\varphi(g)$ can be
found in time $(|g|N)^{O(1)}$ (see (\ref{l1})). Since $f(g)=\psi(\varphi(g))=\psi(k)$ and
$|k|\le|g|$, one can find $\psi(k)$  by Lemma~\ref{ortd} and then to test whether
$g\in\ker(f)$  within the same time. Moreover, due to condition (H3) for cryptosystems
$\S_i$, $i\in\ov n$, one can efficiently find an element $r$ belonging to the right
transversal $R_\varphi$ of $\ker(\varphi)$ in $G$ such that $\varphi(r)=k$ and $|r|\le|k|$.
Now if $g\in\ker(f)$ then $\psi(k)=1_H$ and so $r\in A_\psi$. Furthermore,
$$
\varphi(gr^{-1})=\varphi(g)\varphi(r^{-1})=kk^{-1}=1_K.
$$
Finally, from statement (i3) of Lemma~\ref{spl3} it follows that one can find in time
$(|g|N)^{O(1)}$ an element $a\in A_\varphi$ such that $P_\varphi(a)=gr^{-1}$. Thus
we obtain
$$
P(a,r)=\ov{P_\varphi(a)P_\psi(r)}=\ov{gr^{-1}r}=\ov g=g,
$$
which proves claim~(a).

To prove claim (b) let $g\in G$. If $g\not\in\ker(f)$, then obviously $g\not\in\ker(\varphi)$.
Let now $g\in\ker(f)$ and $(a,b)\in A$ be such that $\ov{P_\varphi(a)P_\psi(b)}=g$.
Since $P_\psi(b)$ belongs to the right transversal $R_\varphi$ of $\ker(\varphi)$ in
$G$, it follows that $g\in\ker(\varphi)$ iff $P_\psi(b)=1_G$. Moreover,
if $P_\psi(b)=1_G$, then obviously $P_\varphi(a)=g$. Taking into account
that the element $P_\psi(b)$ can be found in time $|b|^{O(1)}$ (see~(\ref{ool2})),
we conclude  that the problem $\MM(P_\varphi)$ is polynomial time reducible  to the
problem $\MM(P)$. Thus claim (b) follows from statement (i4) of Lemma~\ref{spl3}.
Theorem~\ref{mres} is proved.\bull

\section{Encrypted simulating of boolean circuits}
\label{sesim}

Let $B=B(X_1,\ldots,X_n)$ be  a boolean circuit and $H$ be a group. Following~\cite{Ba}
we say that a word
\qtnl{di1}
h_1^{X_{l_1}}\cdots h_m^{X_{l_m}},\quad
h_1,\ldots,h_m\in H,\quad l_1,\ldots, l_m\in\ov n,
\eqtn
is a {\it simulation} of size $m$ of $B$ in $H$ if there exists a certain element $h\in H^\#$
such that the equality
$$
h_1^{x_{l_1}}\cdots h_m^{x_{l_m}}=h^{B(x_1,\ldots,x_n)}
$$
holds for any boolean vector $(x_1,\ldots,x_n)\in\{0,1\}^n$.  It is proved in~\cite{Ba} that
given an arbitrary {\it unsolvable} group $H$ and  a boolean circuit~$B$ there exists a
simulation of~$B$ in~$H$, the size of this simulation  is exponential in the depth of~$B$
( in particular, when the depth of $B$ is logarithmic $O(\log n)$, then the size of the
simulation is $n^{O(1)}$).

We say that the circuit $B$ is {\it encrypted simulated} over a homomorphic cryptosystem
with respect to an epimorphism $f:G\to H$ (we use the notations from Definition~\ref{md})
if there exist $g_1,\ldots,g_m\in G$, and a certain element $h\in H^\#$ such that
\qtnl{di2}
f(g_1^{x_{l_1}}\cdots g_m^{x_{l_m}})=h^{B(x_1,\ldots,x_n)}
\eqtn
for any  boolean vector $(x_1,\ldots,x_n)\in\{0,1\}^n$. Thus having a simulation~(\ref{di1})
of the circuit~$B$ in $H$ one can produce an encrypted simulation of $B$ by choosing
randomly $g_i\in G$ such that $f(g_i)=h_i$, $i\in\ov m$ (in this case, equality (\ref{di2})
is obvious). Now combining Theorem~\ref{mres} with the above mentioned result from~\cite{Ba}
we get the following statement.

\crllrl{2602a}
For an arbitrary finite unsolvable group $H$,  a homomorphic cryptosystem $\S$
of a size $N$ over $H$ and  any boolean circuit of the logarithmic depth $O(\log N)$
one can design in time $N^{O(1)}$ an encrypted  simulation of this circuit over $\S$.
\bull
\ecrllr

The meaning of an encrypted simulation is that given (publically) the elements
$g_1,\ldots,g_m\in G$ and $h\in H^\#$ from (\ref{di2}) it should be
supposedly difficult to evaluate $B(x_1,\ldots,x_n)$ since for this purpose one has to
verify whether an element $g_1^{x_{l_1}}\cdots g_m^{x_{l_m}}$ belongs to $\ker(f)$.
On the other hand, the latter can be performed using the trapdoor information.
In conclusion let us mention the following two known protocols of interaction
(cf. e.g. \cite{B,SYY,Ra,Ri}) based on encrypted simulations.

The first protocol is called {\it evaluating an encrypted circuit}. Assume that Alice
knows a trapdoor in a homomorphic cryptosystem over a group $H$ with respect
to an epimorphism $f:G\to H$ and possesses a boolean circuit $B$ which she
prefers to keep secret, and Bob wants to evaluate $B(x)$ at an input
$x=(x_1,\ldots,x_n)$ (without knowing $B$ and without disclosing $x$). To
accomplish this Alice transmits to Bob an encrypted simulation (\ref{di2}) of~$B$,
then Bob calculates the element $g=g_1^{x_{l_1}}\cdots g_m^{x_{l_m}}$ and sends
it back to Alice, who computes and communicates the value $f(g)$ to Bob.
If the depth of the boolean circuit~$B$ is $O(\log N)$ and the homomorphic
cryptosystem is as in Subsection~\ref{mcoac}, then due to
Corollary~\ref{2602a} the protocol can be realized in time $N^{O(1)}$ (here
we make use of that the size of a product of two elements in~$G$ does not
exceed the sum of their sizes).

In a different
setting one could consider in a similar way evaluating an encrypted circuit
$B_H(y_1,\ldots,y_n)$ over a group $H$ (rather than a boolean one), being a
sequence of group operations in $H$ with inputs $y_1,\ldots,y_n\in H$.
The second (dual) protocol is called {\it evaluating at an encrypted input}. Now
Alice has an input $y=(y_1,\ldots,y_n)$ (desiring to conceal it) which she
encrypts randomly by the tuple $z=(z_1,\ldots,z_n)$ belonging to $G^n$
such that $f(z_i)=y_i$, $i\in\ov n$, and transmits $z$ to Bob. In his turn,
Bob who knows a circuit $B_H$ (which he wants to keep secret) yields its
``lifting'' $f^{-1}(B_H)$ to $G$ by means of replacing every constant $h\in H$
occurring in $B_H$ by any $g\in G$ such that $f(g)=h$ and replacing the
group operations in $H$ by the group operations in~$G$, respectively. Then
Bob evaluates the element $(f^{-1}(B_H))(z)\in G$ and sends it back to
Alice, finally Alice applies $f$ and obtains $f((f^{-1}(B_H))(z))=B_H(y)$
(even without revealing it to Bob). Again if the depth of the circuit~$B_H$
is $O(\log N)$ and the homomorphic cryptosystem is as in Subsection~\ref{mcoac},
then the protocol can be realized in time $N^{O(1)}$.
\vspace{4mm}

It would be interesting to design homomorphic cryptosystems over
{\it rings} rather than groups.
\vspace{5mm}

{\bf Acknowledgements.}
The authors would like to thank the Max-Planck Institut fuer Mathematik (Bonn)
during the stay in which this paper was initiated; also Igor Shparlinski for useful
discussions. The research of the second author was supported by grant of
NATO.

\end{document}